%% file: main.tex
\documentclass[sigconf]{acmart}
\renewcommand\footnotetextcopyrightpermission[1]{} 
\usepackage{pgfplots}
\AtBeginDocument{%
  }
\usepackage{subcaption}
\begin{document}

\title{PinLanding: Content-First Keyword Landing Page Generation via Multi-Modal AI for Web-Scale Discovery }

\author{Faye Zhang}  
\email{fzhang@pinterest.com}  
\affiliation{%
  \institution{Pinterest}
  \country{} 
} 
\affiliation{%
  \institution{Stanford University}  
  \country{} 
} 
\orcid{0009-0005-5801-0206}

\author{Jasmine Wan}  
\email{qiwan@andrew.cmu.edu}  
\affiliation{%
  \institution{Carnegie Mellon University}  
  \country{} 
} 

\author{Qianyu Cheng}
\email{qcheng@pinterest.com}
\affiliation{%
  \institution{Pinterest}  
  \country{} 
} 
\orcid{0009-0007-1230-0608}

\author{Jinfeng Rao}
\email{marquisrao@pinterest.com}
\affiliation{%
  \institution{Pinterest}  
  \country{} 
}




\begin{abstract}
  Online platforms like Pinterest hosting vast content collections traditionally rely on manual curation or user-generated search logs to create keyword landing pages (KLPs) --- topic-centered collection pages that serve as entry points for content discovery. While manual curation ensures quality, it doesn't scale to millions of collections, and search log approaches result in limited topic coverage and imprecise content matching. In this paper, we present PinLanding, a novel content-first architecture that transforms the way platforms create topical collections. Instead of deriving topics from user behavior, our system employs a multi-stage pipeline combining vision-language model (VLM) for attribute extraction, large language model (LLM) for topic generation, and a CLIP-based dual-encoder architecture for precise content matching. Our model achieves 99.7\% Recall@10 on Fashion200K benchmark \cite{han2017automatic}, demonstrating strong attribute understanding capabilities. In production deployment for search engine optimization with 4.2 million shopping landing pages, the system achieves a 4X increase in topic coverage and 14.29\% improvement in collection attribute precision over the traditional search log-based approach via human evaluation. The architecture can be generalized beyond search traffic to power various user experiences, including content discovery and recommendations, providing a scalable solution to transform unstructured content into curated topical collections across any content domain.

\end{abstract}

\ccsdesc[500]{Information systems~Information retrieval}
\ccsdesc[500]{Computing methodologies~Natural language processing}
\ccsdesc[300]{Computing methodologies~Computer vision}
\ccsdesc[300]{Computing methodologies~Machine learning}
\ccsdesc[500]{Information systems~Web searching and information discovery}

\keywords{multi-modal AI, content discovery, content organization, search engine optimization, vision-language model, large language model, CLIP, web-scale system, keyword landing page}

\begin{teaserfigure}
  \includegraphics[width=\textwidth]{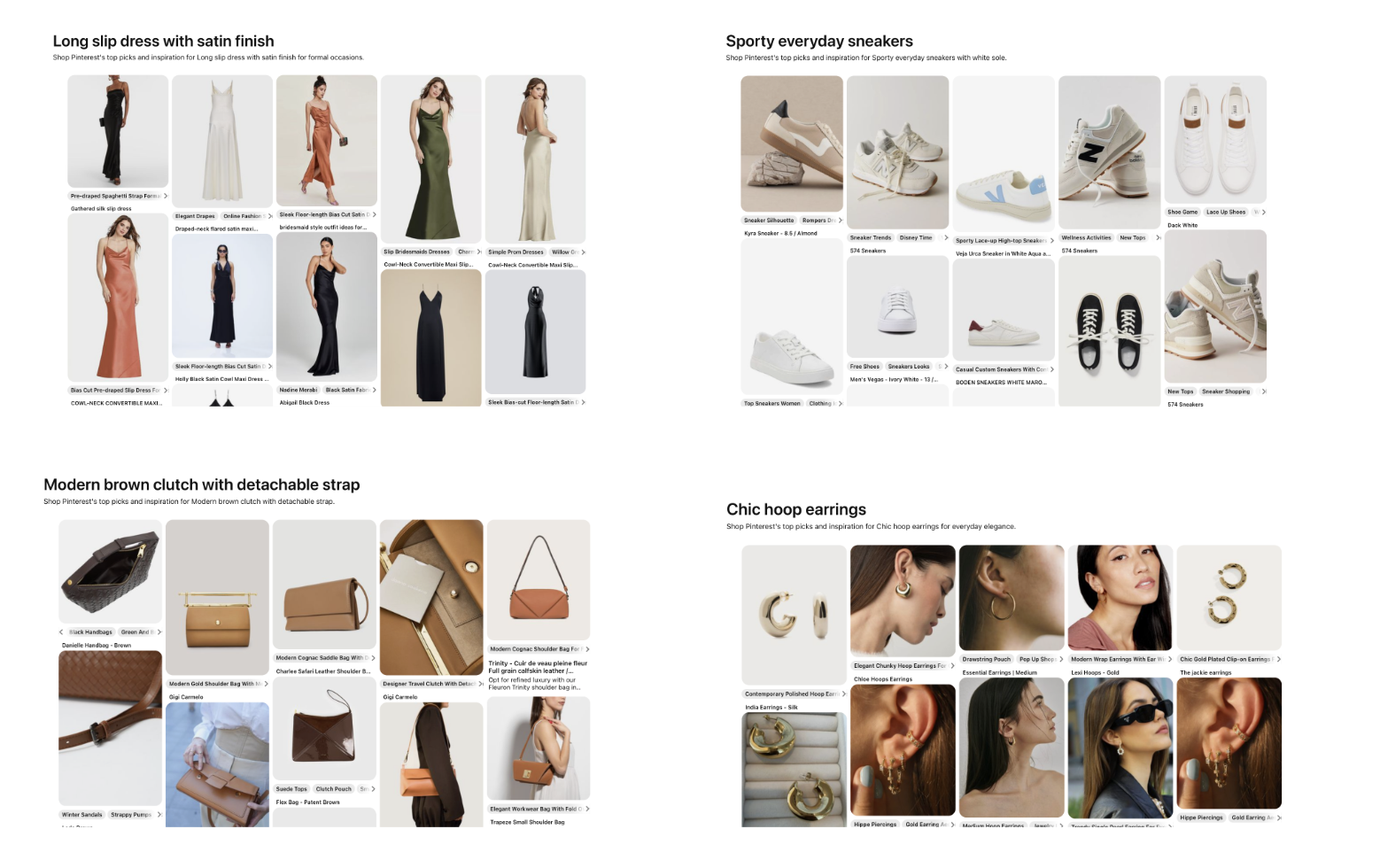}
  \caption{Pinterest Shopping Keyword Landing Pages via PinLanding Architecture, 2025.}
  \label{fig:teaser}
\end{teaserfigure}


\maketitle
\pagestyle{empty}
\input{Sections/introduction}

\input{Sections/relatedwork}

\input{Sections/system_overview}

\input{Sections/experiments_results}

\input{Sections/future_directions}

\input{Sections/conclusion}


\bibliographystyle{ACM-Reference-Format}
\bibliography{bibliography}



\end{document}

%% file: Sections/introduction.tex
\section{Introduction}
Landing pages serve as critical entry points for organic search traffic, enabling platforms to surface their best content to new users. The scale of KLP coverage directly impacts growth potential—platforms with more comprehensive landing pages can capture a broader range of user interests and search intents \cite{seopractice, klp, seooptimization}. Beyond search traffic, well-curated topical collections also provide natural pivot points for users to explore related content and serve as recommendation anchors throughout the browsing journey \cite{topicexplore, topicexplore2}. This makes the ability to generate high-quality topical collections at scale crucial for both acquisition and retention.

Traditionally, the industry standard involves constructing KLPs in two steps \cite{Thomaidou2011MultiwordKR, Cucerzan2007QuerySB}: first, using popular search queries to identify potential landing page topics, then populating each page with top-ranked results from internal search systems for those queries. Major e-commerce platforms \cite{shopifyseo} widely adopt this approach for its simplicity and scalability. However, this approach presents two major limitations: 

\begin{enumerate}
    \item \textbf{Limited Topic Coverage}: Search query-based approaches only capture topics users actively search for, missing valuable content categories in the platform catalog. In our case, this limited us to 1 million landing pages despite having billions of product pins, leaving most content undiscoverable.
    \item \textbf{Precision-Discovery Tradeoff}: While search systems excel at broad content discovery, they often lack the precision \cite{recallprecision} required for curated landing pages. Each piece of content must exactly match specified attributes (e.g., color, brand, product category, material)—a requirement that conflicts with search systems' optimization for broad discovery and exploration.
\end{enumerate}

Recent work has explored deep learning approaches to improve collection curation. Cheng et al. \cite{jie2022deeplearningbasedpage} proposed a two-stage system that first embeds search queries into a semantic space representing purchase intentions, then clusters these embeddings to identify related topics while filtering duplicates. Kouki et al. \cite{collectionretail} developed a system that combines transaction data with product hierarchies to recommend themed collections—for example, identifying complementary home improvement items that are frequently purchased together. While these approaches improved collection quality through better semantic understanding and user behavior analysis, they remain constrained by user search patterns, limiting their ability to discover new topical groupings.

In this paper, we present PinLanding, a novel content-first architecture for generating high-quality topical collections at scale. As shown in Figure  \ref{fig:teaser}, we launched 4.2 million shopping landing pages, demonstrating the practical application of our multi-modal AI framework in enhancing content discovery. Our key contributions are as follows:

\begin{itemize}
    \item We introduce a paradigm shift in collection generation by analyzing content directly rather than relying on user queries, enabling broader coverage and higher precision. 
    \item We develop a scalable multi-modal AI pipeline with two key phases:
    \begin{itemize}
        \item Attribute Generation: Combines VLM for initial attribute extraction with CLIP-based content matching for scalable, consistent attribute assignment.
        \item Collection Creation: Leverages LLM for query generation and distributed processing for precise feed matching via attributes.
    \end{itemize}
    \item We present a highly scalable and cost-efficient architecture that:
    \begin{itemize}
        \item Processes millions of content items with high precision.
       \item Requires no predefined attribute hierarchies or manual curation.
       \item Generalizes across other categories (home decor, beauty and etc.) with minimal adaptation.
    \end{itemize}
\end{itemize}

%% file: Sections/relatedwork.tex
\section{Related Work}
\subsection{Fashion Attribute Generation}
Prior research in fashion attribute generation spans multiple approaches. Multi-task convolutional architectures with fast R-CNN, proposed by Xia et al. \cite{Attributes-oriented}, enabled joint learning of attribute detection and clothing description. Shajini \& Ramanan \cite{9563361} proposed a hierarchical feature fusion framework that combines low-level visual patterns with high-level semantic information through an attention-based semi-supervised approach. Recent work has focused on leveraging large-scale VLMs. FashionVLP \cite{9879706} introduces fashion-specific pretraining for vision-language transformers to capture nuanced attribute relationships. FashionSAP \cite{10204396} demonstrates how fine-grained vision-language pretraining can improve attribute prediction accuracy while maintaining semantic consistency. Our work builds on these advances by introducing a novel pipeline that combines free-form attribute generation from VLM, followed by efficient CLIP-based matching for scalable deployment.

\subsection{Content Organization}
Collection creation requires optimizing both coverage and semantic coherence at scale. Han et al. \cite{8237425} formulated this as a concept discovery problem, introducing spatially-aware methods to automatically learn attribute relationships from visual data. Le et al. \cite{Collaborative} proposed a joint modeling approach that combines preference signals with similarity metrics to enable dynamic collection generation. Our work builds on these directions by introducing a content-first paradigm to create collections that shared the same attributes including visual characteristics, semantic descriptions, and contextual information like occasions.

%% file: Sections/system_overview.tex
\section{Methodology}


\begin{figure}[ht]
    \centering
    \begin{subfigure}[b]{\linewidth}
        \centering
        \includegraphics[width=\linewidth]{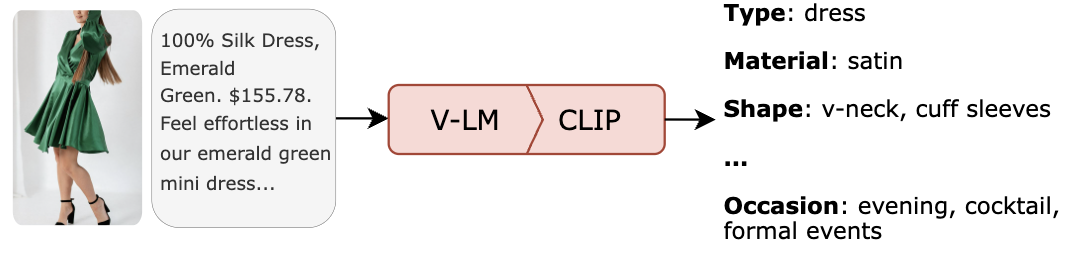}
        \caption{Attribute Generation}
        \vspace{10pt}
    \end{subfigure}

    \begin{subfigure}[b]{\linewidth}
        \centering
        \includegraphics[width=\linewidth]{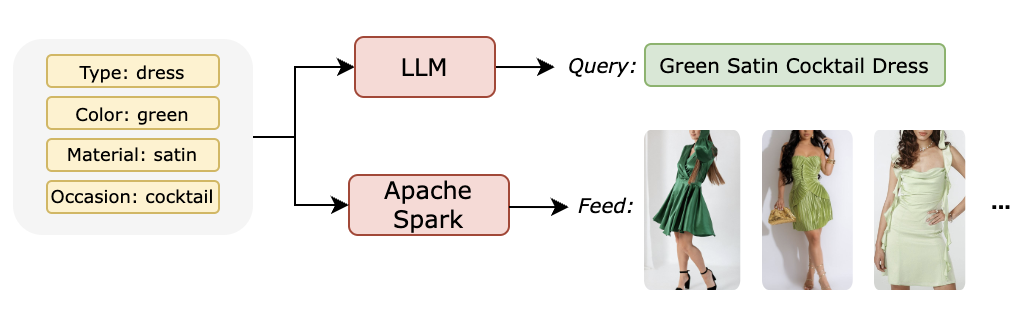}
        \caption{Collection Creation}
    \end{subfigure}
    \caption{Two-Phase Pipeline}
    \label{fig:system}
\end{figure}

We present an end-to-end framework for large-scale topical collection generation that operates in two primary phases: \textbf{attribute generation} and \textbf{collection creation}. Figure \ref{fig:system} illustrates the system architecture.
The first phase addresses the fundamental challenge of extracting structured attributes from the raw content. This phase leverages two interconnected components: Initial Attribute Extraction (Section \ref{sec:initial}) and Content Matching Refinement (Section \ref{sec:content}), ensuring precise and high-quality attribute identification.

The second phase focuses on collection generation through attribute-based topic synthesis. The system selects attribute combinations that are both common and SEO-optimized. Then it proceeds with Query Generation (Section \ref{sec:query}) and Feed Creation (Section \ref{sec:feed}) to support the creation of impactful and engaging KLPs.


\begin{figure*}[ht]
    \centering
    \includegraphics[width=1\linewidth]{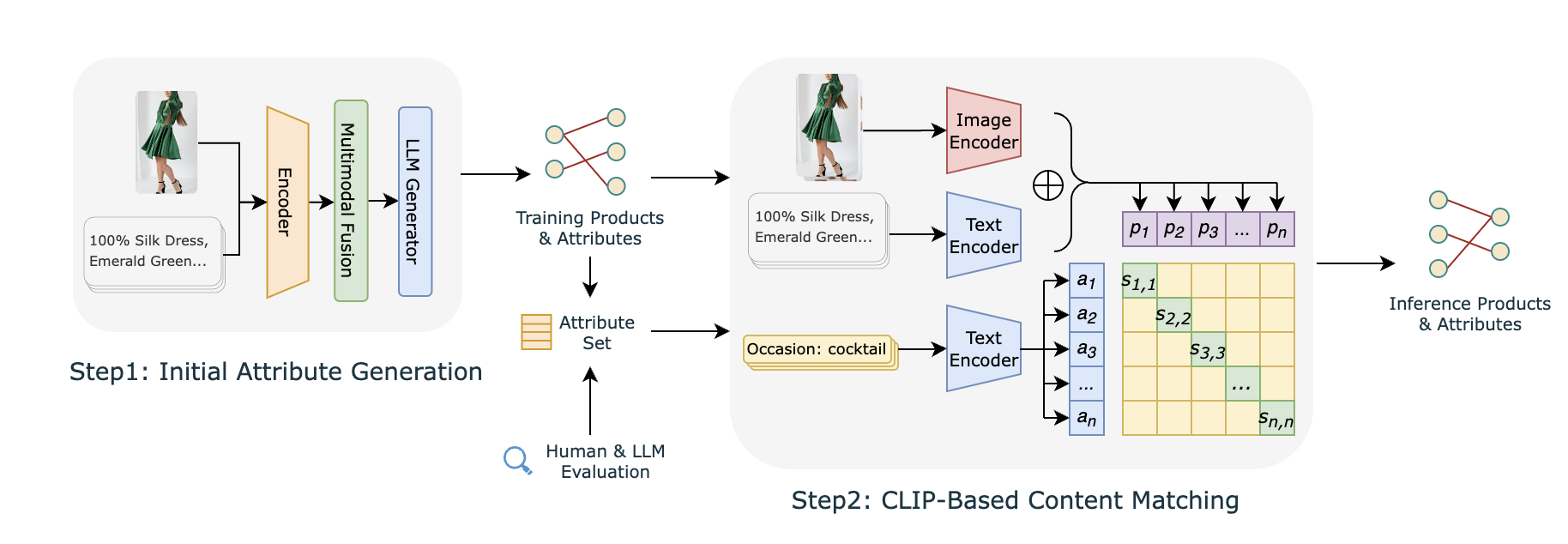}
    \caption{Initial attribute generation and CLIP-based content matching}
    \label{fig:attributes}
\end{figure*}

\subsection{Initial Attribute Extraction}
\label{sec:initial}
As shown in Figure \ref{fig:attributes} (Step 1), we initiate our process by employing the VLM, specifically GPT-4-V \cite{gpt4v, gpt4v2}, which integrates visual capabilities into the GPT-4 \cite{gpt4} framework to generate structured attributes from a diverse set of fashion products. Given a product with image and metadata (title, description, price, merchant-provided tags), we extract attributes across predefined categories including product category, color, style, and occasion. 

Despite the high relevance of the attributes generated by the GPT-4-V, several challenges prevent the direct adoption of these results as the final product attributes:

\begin{enumerate}
     \item \textit{Attribute Quality: } Generated attributes may be overly specific or granular (e.g., "black insoles" for footwear), reducing their utility for collection organization.
    \item \textit{Semantic Consistency:} The model often produces variations of semantically equivalent attributes (e.g., "casual" vs "daywear"), creating redundant annotations.
    \item \textit{Attribute Safety:} Human oversight plays a vital role in identifying and addressing potential biases generated by LLMs, especially those related to gender, race, or culture \cite{llmbias, bias, humanreview}. The free-form generation creates a large volume of unique attributes per product, making manual review for safety and bias infeasible at scale.

\end{enumerate}

\subsection{Content Matching Refinement} 
\label{sec:content}

To achieve scalable attribute assignment while maintaining quality, we develop a refined matching system that processes the initial GPT-4V outputs through attribute curation and model training, as illustrated in Figure \ref{fig:attributes} (Step 2).

\subsubsection{Attribute Set Curation}

We construct a high-quality attribute vocabulary set through a three-stage filtering process:

\begin{enumerate}
    \item Frequency-Based Filtering \\
    Initial attributes are filtered based on occurrence frequency to identify commonly used descriptors. This eliminates rare or overly specific attributes that provide limited value for collection creation \cite{termfrequncy}.

    \item Semantic Deduplication \\
    We compute dense embeddings for each attribute using BERT \cite{bert, bert2} and measure pairwise semantic similarity \cite{dedupe}:
   \[
   \text{sim}(e_i, e_j) = \frac{e_i \cdot e_j}{\|e_i\| \|e_j\|}
   \]
   For attributes with similarity exceeding threshold $\tau$, we retain only the higher-frequency variant, reducing semantic redundancy while preserving coverage.

    \item Human and LLM Quality Review \\
        In collaboration with Pinterest internal AI safety team, both human experts and a LLM systematically review a curated set of approximately 4,000 attributes to ensure their quality and appropriateness. This dual-layer evaluation effectively identifies and mitigates potential biases, particularly in human-related descriptions, while ensuring compliance with product safety standards \cite{dualjudge, llmbias1}. 

\end{enumerate}

\subsubsection{CLIP-based Content Matching}

The initial attributes generated by GPT-4-V, paired with their corresponding products, serve as ground-truth data to train a scalable and effective content matching model.

\paragraph{Mathematical Formalization}  
Consider the set of products \( P = \{p_1, \dots, p_n\} \), where each product \( p_i \) is represented by an image \( x_i \) and a text description \( t_i \). Let \( A = \{a_1, \dots, a_m\} \) denote the set of candidate attributes. The objective of the content matching task is to learn functions \( f \) and \( g \) that map products and attributes into a shared embedding space, such that:  
\[ f: (x, t) \rightarrow \mathbb{R}^d, \quad g: a \rightarrow \mathbb{R}^d, \]  
ensuring that the similarity \( \text{sim}(f(x_i, t_i), g(a_j)) \) is high for correct product-attribute pairs and low for incorrect ones.

\paragraph{CLIP-inspired Model Architecture}  
The model leverages the CLIP architecture \cite{clip}, originally developed for image-text alignment in image captioning, and adapts it for product-attribute matching. As depicted in Figure \ref{fig:attributes} (Step 2), a product is represented by both an image and a text description, and is matched to a corresponding textual attribute.
The product image \( x_i \) is processed through a Vision Transformer (ViT) to generate an image embedding \( f_{\text{img}}(x_i) \). Similarly, the product text \( t_i \) is encoded using a Text Transformer to produce a text embedding \( f_{\text{text}}(t_i) \). These embeddings are combined into a unified product embedding:  
\[
f(x_i, t_i) = f_{\text{img}}(x_i) + f_{\text{text}}(t_i).
\]  
Attributes \( a_i \) are encoded independently using a separate Text Transformer, yielding an attribute embedding \( g(a_i) \). All Transformers (image, product text, and attribute text encoders) are initialized with pretrained weights from CLIP's image and text encoders.

\paragraph{Contrastive Training Objective}

The model is trained using a bi-directional contrastive loss to align matching product-attribute pairs while separating mismatched pairs. The loss function is defined as:  
\[
L = \sum_{i} \left[ -\log \frac{\exp(\text{sim}(f(x_i, t_i), g(a_i^+)) / \tau)}{\sum_{j} \exp(\text{sim}(f(x_i, t_i), g(a_j)) / \tau)} \right],
\]  
where \( a_i^+ \) is the correct attribute for product \( i \), \( \text{sim}(\cdot, \cdot) \) denotes a similarity metric (e.g., cosine similarity), and \( \tau \) is a temperature scaling parameter.
This loss function encourages the model to maximize the similarity between the product embedding \( f(x_i, t_i) \) and its matching attribute embedding \( g(a_i^+) \), while minimizing similarities with non-matching attributes \( g(a_j) \). As a result, the model effectively captures and differentiates relationships between products and attributes.

\paragraph{Post-Processing Weight}

CLIP has been shown to handle long-tailed training data robustly \cite{wenmakes}. However, in our task, this robustness becomes overemphasized—CLIP's predictions overly smooth the natural long-tail distribution of product-attribute relationships in the training data. As a result, common attributes like "color: white" are often underweighted in similarity scores with relevant products, while some rare attributes like "details: embroidered hem" are overweighted.

To address this, we introduce a weight \( w_j \) for each attribute \( j \) based on the frequency of attribute \( j \) in training data, which adjusts the similarity score \( s_{ij} \). The weight is defined as:
\[
w_j = a + b \cdot \text{frequency}_j^{0.5}
\]
where \( a = 1 \) and \( b = 0.01 \) (heuristically chosen). The exponent \( 0.5 \) is used to apply a sublinear scaling to the frequency, ensuring that the adjustment accounts for the magnitude of frequency differences while avoiding an excessive bias toward highly frequent attributes.

The adjusted similarity score is then computed as:
\[
s_{ij}' = w_j \cdot s_{ij} = w_j \cdot \text{sim}(f(x_i, t_i), g(a_j))
\]
This approach ensures that the adjusted predictions better align with the original frequency distribution while retaining the integrity of the similarity computation.

\subsubsection{Product Catalog Attribute Assignment}

During inference, an attribute \( a_j \in A \) is associated with a product \( p_i \in P \) if their similarity surpasses a predefined threshold \( \theta \). This process yields the set of relevant attributes:  
\[
R_i = \{a_j \in A \mid \text{sim}(f(x_i, t_i), g(a_j)) \geq \theta\}.
\]
In this manner, every product in the shopping catalog is assigned a set of relevant attributes.

\subsection{Query Generation}
\label{sec:query}
The query generation phase transforms attribute combinations into natural language queries that serve as collection titles for KLPs \cite{jie2022deeplearningbasedpage}. For example, a combination of attributes (season: summer, color: yellow, type: dress, occasion: party) generates the collection title "Summer yellow dress for parties", providing a natural and searchable entry point for content discovery.

\subsubsection{Attribute Selection}
We formulate attribute combination criteria to ensure collection quality:
\begin{itemize}
   \item Combine 3-4 attributes per query to achieve appropriate specificity
   \item Include at least one category attribute with descriptive attributes
\end{itemize}

\subsubsection{Natural Language Generation}
We utilize GPT-4, a LLM, within a multi-task framework to transform attribute combinations into coherent queries, as illustrated in Table \ref{tab:query-examples}. The LLM serves as both a generator and judge \cite{llmjudge0} to ensure query quality and relevance through three key tasks: 

1. \textit{Semantic Validation:} Assess the semantic validity of attribute combinations to eliminate redundancies, conflicts, or any combinations that result in grammatically incorrect descriptions. (e.g., avoiding "round rings" as rings are inherently round) \cite{llmasjudge, llmjudge1, llmjudge2}

2. \textit{Query Synthesis:} Generate concise, natural queries that align with search patterns (e.g., "Black Long Sleeve Dress for New Year's Eve") \cite{llmquery}

3. \textit{Quality Assessment:} Score generated queries from 1-5 \cite{llmscore} based on anticipated searchability and commonality \cite{common}, filtering out overly specific or unnatural combinations (e.g., "Heavy Wool Sweater for Summer" will be eliminated)

\begin{table}[h]
\caption{Examples of Query Variations Generated from Attributes}
\label{tab:query-examples}
\begin{tabular}{ll}
\toprule
\textbf{Attribute} & \textbf{Generated Query Variations} \\
\midrule
Material: Silk & "Luxury Silk Evening Gowns" \\
             & "Washable Silk Pajama Set" \\
             & "Silk Hair Scrunchies" \\
\midrule
Occasion: Party & "Glamorous Birthday Party Dresses" \\
              & "Garden Party Summer Hat" \\
              & "Halloween Party Costume" \\
\midrule
Style: Casual & "Casual Beach Cover Ups" \\
            & "Everyday Work Backpack" \\
            & "Casual Slip-on Walking Shoes" \\
\bottomrule
\end{tabular}
\end{table}

\subsubsection{Prompt Engineering}
To ensure consistent and high-quality query generation, we develop a comprehensive prompt engineering strategy \cite{prompt, prompt1}. The prompt includes diverse exemplars demonstrating:
\begin{itemize}
   \item Valid and invalid attribute combinations
   \item Quality scoring criteria with examples spanning the full range (1-5)
   \item Natural language variations for different attribute combinations
\end{itemize}

These examples guide the model in maintaining semantic validity while producing natural language queries suitable for collection titles.

\subsubsection{Quality Control}
We implement a multi-stage filtering process for the generated queries:
\begin{itemize}
   \item Semantic validation: Remove queries with invalid grammar or attribute combinations
   \item Searchability: Retain only queries with searchability scores $\geq$ 4
\end{itemize}

This systematic filtering ensures that only high-quality, semantically valid queries with adequate product coverage are used for collection creation.

\subsection{Feed Generation}
\label{sec:feed}
The final phase of our pipeline focuses on efficiently creating product collection for each query. Given millions of products and queries, this requires careful optimization of distributed processing \cite{spark}.

\subsubsection{Matching Algorithm}
We implement an attribute-based matching system using Apache Spark for distributed computation \cite{Ko2016ProcessingLD}. For each query-product pair, we calculate a relevance score based on attribute overlap:

\[
\text{relevance}(q, p) = \sum_{a \in A_q \cap A_p} \text{score}(a)
\]

where $A_q$ and $A_p$ are the attribute sets of query q and product p respectively, and score(a) represents the confidence score of attribute a.

\subsubsection{Matching Algorithm}
We implement an attribute-based matching system using Apache Spark for distributed computation \cite{Ko2016ProcessingLD}. For each query-product pair, we calculate a relevance score based on attribute overlap:

\[
\text{relevance}(q, p) = \sum_{a \in A_q \cap A_p} \text{score}(a)
\]

where $A_q$ and $A_p$ are the attribute sets of query q and product p respectively. The attribute confidence score $\text{score}(a)$ is computed as:

\[
\text{score}(a) = \text{sim}(f(p), g(a)) \cdot w(a)
\]

Here, $\text{sim}(f(p), g(a))$ is the CLIP similarity between product p and attribute a, and $w(a)$ is a popularity-based weight calculated as:

\[
w(a) = 1 + \sqrt{\text{freq}(a)}/100
\]

where $\text{freq}(a)$ is the frequency of attribute a in our training data. This weighting scheme helps balance between attribute precision and coverage.

\subsubsection{Computational Optimizations}
To process large-scale matching efficiently, we implement several key optimizations:

1. Attribute Score Caching:
  \begin{itemize}
      \item Pre-compute a mapping of attributes to confidence scores
      \item Store in distributed cache for O(1) lookup complexity
      \item Eliminate redundant score computations
  \end{itemize}

2. Batch Processing:
  \begin{itemize}
      \item Group products by shared attributes to reduce computations
      \item Pre-filter candidates using attribute occurrence thresholds
      \item Optimize join operations through data partitioning
  \end{itemize}

3. Memory Management:
  \begin{itemize}
      \item Cache frequently accessed attribute mappings
      \item Implement efficient data structures for attribute lookup
      \item Optimize shuffle operations in distributed processing
  \end{itemize}

These optimizations reduced processing time by 92\% while maintaining matching precision. The system ensures collection quality by requiring a minimum of 20 products per collection, enforcing relevance score thresholds, and validating product trust criteria.

%% file: Sections/experiments_results.tex
\section{Experiments and Results}

\subsection{Implementation Details}
\subsubsection{Dataset Composition}
The training dataset comprises 200,000 fashion products across apparel, accessories, and footwear categories. Each product is characterized by multimodal content, including product images and metadata such as title, description, price, and tags provided by merchants. Furthermore, attributes are generated via VLM, encompassing a comprehensive three-level category hierarchy. This hierarchy includes broad categories (e.g., Apparel), subcategories (e.g., Dresses), and specific types (e.g., Maxi Dresses). Descriptive attributes are defined to capture key product features, including color, primary material, fit, stretch, shape, style, gender, age group, price level, season, festival, occasion, and brand.


\subsubsection{Training Implementation Details} Our model extends CLIP-H/14 architecture (1.3B parameters) \cite{Li2023CLIPAv2SC} with a dual-encoder architecture designed for multimodal representation learning. The text encoder comprises 24 transformer layers with a hidden dimension of 1,024, utilizing a token embedding vocabulary of 32,100 tokens and a maximum sequence length of 77 tokens. The vision encoder features 32 transformer layers with a hidden dimension of 1,280, configured with a 14×14 patch embedding and stride of 14. 

Training is optimized using FusedAdam \cite{apex2020nvidia}, an efficient implementation of the Adam optimizer that reduces memory bandwidth through operation fusion. The learning rate is set to $5 \times 10^{-8}$ with cosine annealing schedule.

\subsection{Experimental Results}
\subsubsection{Evaluation on Fashion200K Benchmark} We evaluate our \\
model's attribute prediction accuracy on the Fashion200K dataset \cite{han2017automatic}, a standard benchmark for fashion understanding tasks. The dataset contains fashion product images with hierarchical category labels spanning major categories (dresses, pants, tops) and subcategories (e.g., casual dresses, formal dresses).

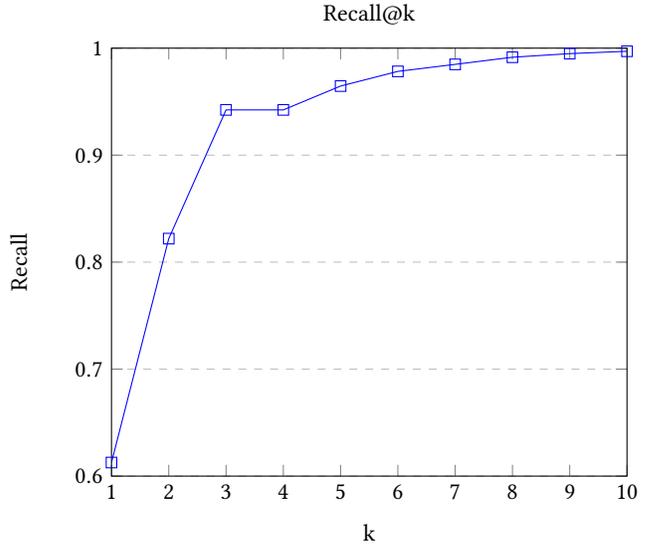
\begin{figure}[H]
    \centering  
    \begin{tikzpicture}  
        \begin{axis}[  
            title={Recall@k},  
            xlabel={k},  
            ylabel={Recall},  
            xmin=1, xmax=10,  
            ymin=0.6, ymax=1,  
            xtick={1,2,3,4,5,6,7,8,9,10},  
            ytick={0.6,0.7,0.8,0.9,1.0},  
            legend pos=north east,  
            ymajorgrids=true,  
            grid style=dashed,  
        ]  
        \addplot[  
            color=blue,  
            mark=square,  
            ]  
            coordinates {  
            (1,0.6126873661670236)(2,0.8219686830835118)(3,0.9422845289079229)(4,0.9422845289079229)(5,0.9645677194860813)(6,0.9782855995717344)(7,0.9848434154175589)(8,0.9915016059957173)(9,0.9948808886509636)(10,0.9970891327623126)  
            };  
        \end{axis}  
    \end{tikzpicture}  
    \caption{Recall@k for k from 1 to 10 on Fashion200K dataset}  
    \label{tab:recall_chart}  
\end{figure}

We measure performance using Recall@k metrics, which assess whether the true category label appears among the model's top k predicted attributes. Our model achieves \textbf{99.7\% }Recall@10 on Fashion200K compared to other benchmarks in Table \ref{tab:benchmarks}. Figure \ref{tab:recall_chart} shows detailed performance across different k values. This strong performance on a public benchmark validates our attribute extraction approach and demonstrates that our training strategy, which leverages a diverse fashion product catalog, enables robust fashion attribute understanding.

\begin{table}[h]
    \centering
    \caption{Recall@10 on the Fashion200k dataset}
    \begin{tabular}{lc}
        \toprule
        Method & R@10 \\
        \midrule
        RN \cite{NIPS2017_e6acf4b0} & 40.5 \\
        MRN \cite{mrn} & 40.0 \\
        TIRG \cite{trig} & 42.5 \\
        CosMo \cite{cosmo} & 50.4 \\
        FashionVLP \cite{9879706} & 49.9 \\
        VAL \cite{val} & 53.8 \\
        LLM-MS \cite{vlm} & 71.4 \\
        \textbf{PinLanding} & \textbf{99.7} \\
        \bottomrule
    \end{tabular}
    \label{tab:benchmarks}
\end{table}

\subsubsection{Collection Quality Analysis}
Table \ref{tab:precision_table} shows human-evaluated precision@10 comparisons between collections generated by our approach versus the search-based baseline. Overall, our method achieves a \textbf{14.29\%} improvement in collection precision across all categories. The evaluation measures whether products in each collection accurately reflect their collection title (e.g., "Blue Silk Dresses for Summer Parties" should contain dresses that are blue, made of silk, and suitable for summer parties).

\begin{table}[h]  
    \centering  
    \caption{Precision@10 Comparison for Collection Quality}  
    \begin{tabular}{@{}lcc@{}}  
        \toprule  
        Category & Search Log & PinLanding \\ \midrule  
        Color & 0.88 & \textbf{0.97} \\  
        Main Material & 0.77 & \textbf{0.96} \\  
        Fit \& Stretch & 0.90 & \textbf{0.96} \\  
        Shape & 0.84 & \textbf{0.97} \\  
        Details & 0.87 & \textbf{0.94} \\  
        Style & 0.88 & \textbf{1.00} \\  
        Season \& Festival & 0.77 & \textbf{0.94} \\  
        Occasion & 0.84 & \textbf{0.87} \\  
        Brand & 0.77 & \textbf{1.00} \\ \midrule
        Average & 0.84 & \textbf{0.96} \\ \bottomrule  
    \end{tabular}  
    \label{tab:precision_table}  
\end{table}

\subsubsection{Production Impact}
The framework has been deployed in two key applications:

1. Shopping KLP Generation for Search Traffic Optimization:
   \begin{itemize}
       \item Generation of 4.2 million fashion landing pages, representing a 4X increase in unique topics over the traditional user search log based approach
       \item +35\%  search engine index rate \cite{index} compared to Pinterest baseline  
   \end{itemize}

2. Related Content Discovery:
We extend the framework to enable semantic navigation between collections by leveraging the learned topic embeddings. The system computes similarity between collection embeddings to identify semantically related content, enabling users to explore related collections based on attribute relationships \cite{relatedcontent}. This semantic navigation layer improves content discovery by surfacing thematically related collections during user browsing sessions.

\subsection{System Analysis}

\subsubsection{Computational Requirements}
The system demonstrates practical computational requirements for enterprise deployment. Training completes in 12 hours using 8 NVIDIA A100 GPUs, with a total estimated cost of approximately \$500 per training run. 

\subsubsection{Domain Adaptability}
Our framework exhibits domain-agnostic properties through three key adaptation mechanisms:
\begin{itemize}
   \item Attribute Extraction: Domain-specific characteristics are captured through prompt refinement of the VLM while maintaining the base neural architecture
   \item Topic Synthesis: Flexible attribute-to-topic generation enables diverse collection styles, from functional ("Modern Living Room Furniture") to inspirational ("Coastal Living Room Ideas")
   \item Matching Alignment: Lightweight fine-tuning of the CLIP-based model enables consistent performance across domains
\end{itemize}

\subsubsection{Broader Applications}
While our primary experiments focus on search-optimized collections, the content-first paradigm enables several additional applications:

1. Automated Merchandising: The framework enables systematic generation of thematic collections by:
  \begin{itemize}
      \item Learning temporal patterns in attribute relationships
      \item Identifying emerging style combinations
      \item Generating collections that align with merchandising strategies
  \end{itemize}

2. Semantic Discovery: The learned attributes support exploration-focused content organization through:
  \begin{itemize}
      \item Hierarchical organization of related collections
      \item Dynamic generation of contextual navigation paths
      \item Attribute-aware content recommendations
  \end{itemize}

%% file: Sections/future_directions.tex
\section{Future Directions}

While our content-first approach demonstrates significant improvements in collection generation, several open challenges and research opportunities remain. Our current framework, while effective for attribute-based collections, faces limitations in capturing emerging cultural phenomena that manifest in search behavior \cite{Kralisch2004CulturalDO, googletrend}. 2024 trending fashion queries like "Old Money Aesthetic" or "Barbiecore" \cite{barbircore} represent complex style concepts that emerge organically from social discourse rather than explicit attribute combinations.

To address these challenges, we propose exploring an AI agent-based framework for automated collection orchestration \cite{agent, agent2}. This presents several key research directions:
\begin{itemize}
    \item Semantic Decomposition: Methods for decomposing abstract style concepts into concrete visual and contextual attributes
    \item Cross-Modal Alignment: Approaches for aligning social discourse with visual content representations
    \item Agent-based Orchestration: Developing AI agents that can monitor trends, understand emerging style concepts, and dynamically adjust collection generation parameters
\end{itemize}

Such an agent-based system could enable real-time adaptation of our collection generation process while maintaining the precision of our attribute-based approach. This represents a step toward more dynamic and culturally-aware content organization systems that combine bottom-up attribute analysis with top-down trend understanding.

%% file: Sections/conclusion.tex
\section{Conclusion}

This paper presents a novel content-first approach for large-scale topical collection generation using multimodal AI. Our framework leverages VLM for attribute extraction, LLM for topic generation, and efficient CLIP-based matching for scalable deployment. The system's effectiveness is validated through large-scale production deployment, demonstrating significant improvements over traditional search log based approaches.

Our architecture's domain-agnostic properties enable extension to various categories with minimal modifications. While the current framework excels at attribute-based collection generation, future work will explore AI agent-based approaches for capturing emerging style concepts and cultural trends. This represents a step toward more dynamic content organization systems that can combine structured attribute analysis with evolving user interests.

Our work contributes to the broader research agenda of scalable content organization and demonstrates how modern AI architectures can enable more comprehensive and precise collection generation.